%% file: main.tex
\documentclass[conference]{IEEEtran}
\IEEEoverridecommandlockouts
\usepackage{cite}
\usepackage{amsmath,amssymb,amsfonts,bm}
\usepackage{algorithm}
\usepackage{algpseudocode}
\usepackage{graphicx}
\usepackage{textcomp}
\usepackage{xcolor}
\usepackage{siunitx}
\usepackage{booktabs}
\usepackage{xspace}
\usepackage{soul}
\usepackage{eso-pic}
\usepackage[dvipsnames]{xcolor}
\usepackage{upgreek}
\usepackage{enumitem}
\usepackage[most]{tcolorbox}
\usepackage{multirow}
\usepackage{makecell}
\usepackage{cuted}
\usepackage{fancyhdr}

\usepackage{booktabs}
\usepackage{float} 
\usepackage{tikz}
\def\BibTeX{{\rm B\kern-.05em{\sc i\kern-.025em b}\kern-.08em
    T\kern-.1667em\lower.7ex\hbox{E}\kern-.125emX}}

\input{macros}

\fancypagestyle{ieeecopyright}{
  \fancyhf{}
  
  \fancyfoot[C]{\parbox{\textwidth}{\centering \fontsize{7}{8}\selectfont \textcopyright~2026 IEEE. Personal use of this material is permitted. Permission from IEEE must be obtained for all other uses, including reprinting/republishing this material for advertising or promotional purposes, collecting new collected works for resale or redistribution to servers or lists, or reuse of any copyrighted component of this work in other works.}}
}

\begin{document}

\title{{\system}: Arbitrary RF Containment Using Tactical Aerial Networks and Differentiable Ray Tracing}

\author{%
\IEEEauthorblockN{Samuel Rivera, Zhihui Gao, Yiming Li, Tingjun Chen}
\IEEEauthorblockA{
Department of Electrical and Computer Engineering, Duke University Durham, NC, USA\\
\{sam.rivera, zhihui.gao, yiming.li, tingjun.chen\}@duke.edu}
}

\maketitle

\thispagestyle{ieeecopyright}

\input{sections/abstract}

\input{sections/intro}

\input{sections/related_works}

\input{sections/system_model}

\input{sections/problem_formulation}

\input{sections/experiments}

\input{sections/conclusion}
\input{sections/acknowledgements}

\bibliographystyle{IEEEtran}
\bibliography{reference}

\end{document}

%% file: macros.tex
\definecolor{brightgreen}{rgb}{0.4, 1.0, 0.0}

\DeclareSIUnit{\belmilliwatt}{Bm}
\DeclareSIUnit{\dBm}{\deci\belmilliwatt}

\newcommand{\myparatight}[1]{\vspace{0.75ex}\noindent\textbf{#1~~}}

\newcommand{\txset}{\mathcal{B}}         %

\newcommand{\nodeidx}{n}                 %
\newcommand{\nodeset}{\mathcal{N}}       %
\newcommand{\nodenum}{N}

\newcommand{\jamset}{\mathcal{J}}        %

\newcommand{\cjacro}{\text{CJ}}

\newcommand{\shapes}{\text{Circle/L-Shape/Square}}

\newcommand{\areaTotal}{\mathcal{A}}
\newcommand{\areaBuilding}{\areaTotal_{0}}

\newcommand{\areaTarget}{\mathcal{Z}}
\newcommand{\zoneInside}{\mathcal{Z}_{\text{in}}}
\newcommand{\zoneOutside}{\mathcal{Z}_{\text{out}}}

\newcommand{\RT}{\text{RT}}
\newcommand{\pidx}{p}

\newcommand{\x}[1]{x_{#1}}
\newcommand{\y}[1]{y_{#1}}
\newcommand{\z}[1]{z_{#1}}
\newcommand{\az}[1]{\alpha_{#1}}
\newcommand{\el}[1]{\beta_{#1}}
\newcommand{\powerTx}[1]{P^{\text{tx}}_{#1}}

\newcommand{\rxSet}{\mathcal{M}}
\newcommand{\rxSetInside}{\mathcal{M}_{\text{in}}}
\newcommand{\rxSetOutside}{\mathcal{M}_{\text{out}}}

\newcommand{\powerRx}[1]{P_{#1}}
\newcommand{\powerNoise}{P_{0}}
\newcommand{\RX}{\text{RX}}

\newcommand{\rxidx}{m}
\newcommand{\rx}[1]{\mathbf{r}_{#1}}

\newcommand{\ABS}{\text{Aerial Base Stations}}             %

\newcommand{\txConfig}{\bm{\uptheta}}
\newcommand{\txConfigSet}{\bm{\Theta}}

\newcommand{\SINR}{\Gamma}
\newcommand{\SINRlin}{\SINR_m}          %
\newcommand{\powerTxMin}{P^{\text{tx}}_{\text{min}}}
\newcommand{\powerTxMax}{P^{\text{tx}}_{\text{max}}}
\newcommand{\Sin}{\rx{\mathrm{in}}}     %

\newcommand{\Loss}{\mathcal{L}}

\newcommand{\Lin}{\mathcal{L}_{\mathrm{in}}}   %
\newcommand{\Lout}{\mathcal{L}_{\mathrm{out}}} %
\newcommand{\Lunif}{\mathcal{L}_{\mathrm{u}}}  %
\newcommand{\Lgate}{\mathcal{L}_{\mathrm{g}}}  %

\newcommand{\wIn}{\lambda_{\text{in}}}
\newcommand{\wOut}{\lambda_{\text{out}}}
\newcommand{\wUnif}{\lambda_{\text{u}}}
\newcommand{\wGate}{\lambda_{\text{g}}}

\newcommand{\SINRthresh}{\gamma}             %
\newcommand{\SINRfloor}{\eta}               %
\newcommand{\splus}{\varphi}                 %
\newcommand{\muin}{\mu_{\text{in}}}     %
\newcommand{\muout}{\mu_{\text{out}}}

\newcommand{\elmin}{\beta_{\min}}
\newcommand{\elmax}{\beta_{\max}}

\newcommand{\freq}{f_c} %

\newcommand{\system}{ARCTAN}

\newcommand{\iterstage}{I}
\newcommand{\iter}[1]{I_{#1}}

%% file: sections/abstract.tex
\begin{abstract}
Aerial base stations (ABSs) can rapidly establish connectivity in ad hoc, infrastructure-deprived environments, but their broadcast, line-of-sight transmissions leak far beyond the intended service area, exposing communications to passive eavesdropping and interference.
Prior physical layer defenses based on cooperative jamming typically assume known eavesdropper locations, simplified statistical channels, or continuously repositioned jammers. 
We instead pose the problem as a radio frequency (RF) containment: confining usable signal to a user-defined, arbitrarily-shaped target zone while denying it elsewhere independent of eavesdropper location.
We present {\system}, a gradient-based optimization framework that jointly optimizes the position, orientation, and transmit power of stationary ABSs and cooperative jammers (CJs) by backpropagating through site-specific, differentiable 3D ray traced channels.
Evaluated in a high-fidelity digital twin across three target zone geometries, {\system} achieves a mean in-zone SINR of approximately 10\,dB while reducing mean out-of-zone SINR from {13-16}\thinspace{dB} to $-${3--5}\thinspace{dB}, and suppressing signal-leakage ratios from over 93\% to below 46\% requiring at most 10 of 12 candidate CJs.
\end{abstract}

%% file: sections/intro.tex
\section{Introduction and Motivation}
\label{sec:intro}

Military tactical networks increasingly depend on rapidly deployable, ad hoc topologies to sustain operations in contested and infrastructure-deprived environments, where fixed terrestrial infrastructure may be degraded, denied, or altogether absent. An emerging approach for establishing on-demand connectivity for a given \emph{target zone} is the deployment of aerial base stations (ABSs), which offer near immediate access to line-of-sight channels and high capacity coverage~\cite{rolly2022unmanned}.
Nevertheless, the broadcast nature of ABS transmissions allows signals to propagate far beyond their operational area, making communication susceptible to passive eavesdropping from terrestrial adversaries~\cite{tippenhauer2013limitations}. Prominent work utilizes physical layer security (PLS), or more specifically \emph{cooperative jammers (CJs)}, whose interference degrades the signal-to-interference-plus-noise ratio (SINR) observed by an eavesdropper while limiting disruptions in communication between legitimate nodes. While effective in principle, solutions often presume a known or bounded adversary location~\cite{lee2018uav, diao2022enhancing, roh2019cooperative}, use simplified channel models~\cite{kim2021securing, jiang2025friendly}, and/or rely on the continuous repositioning of CJs that are typically resource-constrained~\cite{zhong2018secure, li2018mobile, lei2024multi}.

To address each limitation in full, we refactor aerial ad hoc coverage design as an \emph{RF containment problem: confining the usable signal to a user-defined, arbitrarily shaped target zone while denying it everywhere else}, regardless of an eavesdropper's physical location. We use the phrase ``RF containment'' in an operational rather than physical sense, where the goal is not to attenuate the radiated field beyond the target zone, but to deny the usable signal there, thereby driving the SINR below the decoding threshold of a conventional receiver. Realizing this containment in a real deployment requires jointly reasoning over the placement, orientation, and power of every airborne node in the operational environment.

We translate this strategy into a functional architecture known as {\system}, or \underline{\textbf{A}}rbitrary \underline{\textbf{R}}F \underline{\textbf{C}}ontainment using \underline{\textbf{T}}actical \underline{\textbf{A}}erial \underline{\textbf{N}}etworks, a dual-objective optimization framework that aims to maximize and smooth the SINR inside a specified target zone while minimizing the SINR outside it. This is achieved via the joint optimization of the position, orientation, and transmit power of a given set of ABSs and CJs. Rather than relying on statistical channel approximations, this framework utilizes \emph{differentiable ray tracing (RT)} so that gradient-based methods can explicitly tailor coverage to the physical geometry of the environment~\cite{sionnarttech}. Furthermore, {\system} effectively minimizes the overall deployment cost through learnable activation gates that select the minimally viable set of CJs under sustained operation, limiting excess RF interference and co-altitude crowding in shared airspace.

To evaluate {\system}, we use Sionna RT~\cite{sionnarttech} coupled with a high-fidelity digital twin framework offered by Geo2SigMap~\cite{li2024geo2sigmaphighfidelityrfsignal}, as illustrated in Fig.~\ref{fig:zones}. We demonstrate its generality using three representative target zones denoted by {\shapes}. Results show that {\system}'s gradient-based optimization successfully confines ABS coverage to each of the designated target zones with a mean SINR of {10.0/9.7/9.6}\thinspace{dB}.
With the assistance of optimally placed CJs, {\system} is shown to limit signal ``leakage'' from {94.7}\%, {94.9}\%, and {93.2}\% to {43.9}\%, {42.4}\%, and {45.2}\%.

In summary, the contributions of this paper include:
\begin{itemize}[leftmargin=*, topsep=2pt, itemsep=1pt]
    \item
    A unique formulation for RF coverage containment as a spatial SINR optimization over arbitrarily shaped target zones, with a dual-purpose goal of maximizing and minimizing the SINR inside and outside the zone, respectively;
    \item
    {\system}, a gradient-based optimization framework that jointly solves for the configurations of both ABSs and CJs to satisfy its given objective;
    \item
    An evaluation of {\system} in a digital twin environment using Sionna RT, demonstrating its %
    effectiveness for several unique zones in a single operational environment.
\end{itemize}

%% file: sections/related_works.tex
\section{Related Works}
\label{sec:related_works}

PLS via cooperative jamming provides a robust alternative to cryptographic countermeasures, which impose computational overhead unsuitable for low-power embedded platforms and remain vulnerable to advances in cyber warfare~\cite{illi2023physical}. Among prior work on countermeasures against terrestrial passive eavesdroppers, three limiting assumptions are prevalent. First, most approaches presume that channel state information (CSI) and adversary locations are known, partially-known or boundable in advance~\cite{lee2018uav, kim2021securing, roh2019cooperative, jiang2025friendly, wen2024covert, liu2026friendly}. While this is plausible against an actively transmitting eavesdropper, tactical settings frequently encounter passive adversaries that emit no discernible channel information for localization. Second, existing architectures achieve robustness by continuously repositioning CJs to track a mobile or emerging threat, incurring energy overheads incompatible with the power budgets of low size, weight, and power (SWaP) platforms. Moreover, system performance is characterized using statistical channel models (e.g., Nakagami-$m$) that do not capture the multipath-rich geometry of urban or heavily obstructed propagation environments~\cite{kim2021securing, jiang2025friendly}. Third, some frameworks pursue low-probability-of-detection (LPD) covertness aided by CJs~\cite{chen2023achieving, liu2026friendly, wen2024covert} which trade concealment for the sustained, high-throughput communication required by tactical deployments.

In parallel, a separate body of work removes the need for adversary localization altogether by reasoning about security spatially: rather than nulling a specific eavesdropper in space, these methods define secrecy-coverage metrics over a bounded secrecy region, a notion first introduced in~\cite{vilela2011secrecy} and extended in~\cite{zhou2018improving, zhang2023achieving, cabezas2021weighted}. This framing is conceptually closest to our own, but its execution remains limited in comparison. Existing secrecy-region analyses assume rudimentary coverage zones and use simplified LOS channel geometries, making their adoption in realistic propagation environments infeasible. In contrast, this gap has been more thoroughly explored in the study of capacity and coverage optimization (CCO) for cellular networks, where ray-traced channels are used to optimize %
the system parameters of both
terrestrial~\cite{yildiz2025digital, liu2019efficient} and non-terrestrial~\cite{belgiovine2025better, cho2025optimal, yoon2025study} base stations. 

{\system} bridges these two lines of work, importing the site-specific precision of ray-traced CCO into the tactical PLS problem while reframing the objective from covertness (LPD) to low-probability-of-intercept (LPI): denying a transmission's usability beyond a target zone, rather than concealing its existence. Prior spatial PLS work fixes this boundary analytically, as a function of the channel model and assumed adversary class. {\system} instead treats the boundary as an operator-specified input, formulating containment as a gradient-based optimization that backpropagates directly through differentiable, site-specific ray-traced channels. To our knowledge, {\system} is the first framework to jointly support multi-objective coverage optimization while accommodating arbitrarily-shaped, operator-defined target zones using site-specific RT-based channel models.

%% file: sections/system_model.tex
\section{System Model}
\label{sec:system}

\begin{figure*}[!t]
    \centering
    \includegraphics[width=1.0\textwidth]{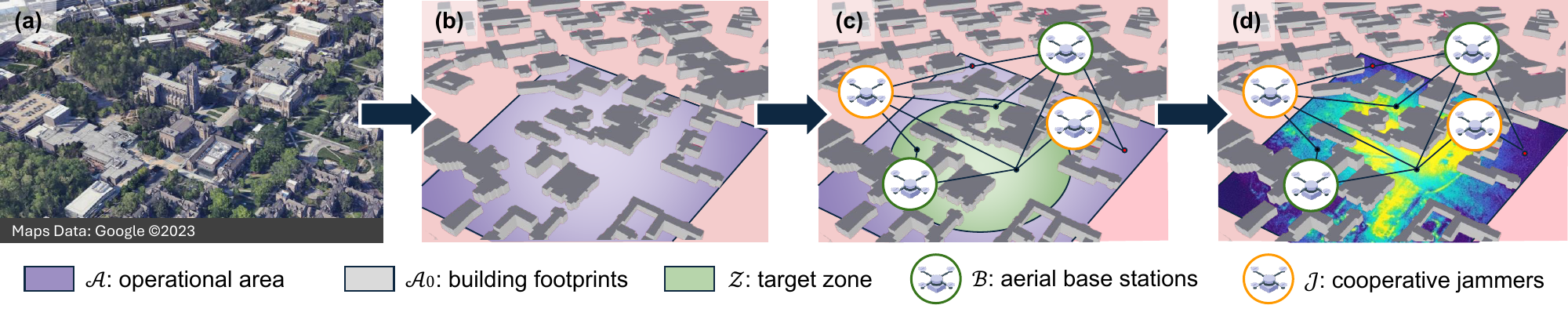}
    \vspace{-3mm}
    \caption{Walkthrough of {\system}:
    (a) Real-world view of the Duke Campus quad;
    (b) 3D scene generated by Geo2SigMap with the operational area $\areaTotal$;\
    (c) Formation of Target Zone $\areaTarget$ and initialization of both ABSs and CJs;
    (d) Optimized spatial coverage with {\system}.}
    \label{fig:zones}
    \vspace{-3mm}
\end{figure*}

\subsection{Network Model and Node Roles}
We consider a low-altitude flying ad-hoc network (FAHN) deployed in a 3D Cartesian coordinate system.
The airborne transmitters (TXs) are partitioned by their role and capability into a set of service \emph{aerial base stations (ABSs)}, $\mathcal{B}$, and a set of co-channel \emph{``cooperative'' jammers (CJs)}, $\mathcal{J}$, operating at frequency $\freq$=$\text{3.7 GHz}$.
We use $\mathcal{N} \triangleq \mathcal{B} \cup \mathcal{J}$ to denote the set of airborne nodes, with $|\nodeset| = |\txset| + |\jamset| = \nodenum$.

Each node $\nodeidx \in \nodeset$ is modeled by a steerable point source, whose configuration is given by
\begin{equation}
  \txConfig_{\nodeidx} = \left( \x{\nodeidx}, \y{\nodeidx}, \z{\nodeidx}, \az{\nodeidx}, \el{\nodeidx}, \powerTx{\nodeidx} \right),\ \nodeidx \in \nodeset,
  \label{eq:theta}
\end{equation}
where $(\x{\nodeidx}, \y{\nodeidx}, \z{\nodeidx})$ denotes the node location in space, $\az{\nodeidx}$ and $\el{\nodeidx}$ are the azimuth and elevation steering angles, and $\powerTx{\nodeidx}$ is the TX power.
The ABSs and CJs share this parameterization, but differ in the antenna patterns:
The ABSs utilize directional antennas with the 3GPP TR 38.901 radiation pattern specifications~\cite{3gpp.38.901}, while the CJs use isotropic antennas as the broad-area interferers.

\subsection{Coverage Zones and Sampled Receiver ($\RX$) Locations}
\label{sub:coveragezones}

Consider a bounded \emph{operational area} $\areaTotal \subseteq \mathbb{R}^{2}$ (Figs.~\ref{fig:zones}(a--b)).
In contrast to conventional BS sectorization, the ABSs jointly serve an outdoor, \emph{arbitrarily shaped target zone}, denoted by $\areaTarget \subseteq \areaTotal$ (Fig.~\ref{fig:zones}(c)).
We treat $\areaTarget$ abstractly as any bounded region delimited by an operator-defined or mission-specific boundary $\partial\areaTarget$, which may overlap with buildings.
Let $\areaBuilding \subseteq \areaTotal$ denote the union of all building footprints.
As we only consider outdoor operational scenarios, the buildings are masked out, and the remaining outdoor area $\areaTotal \setminus \areaBuilding$ is partitioned into an interior zone $\zoneInside$ and an exterior zone $\zoneOutside$.

We evaluate the SINR at a set of RX samples indexed by $\rxSet = \{1, \dots, M\}$.
The $\rxidx$-th RX sample is a 3D point on the receiver plane, given by $\rx{\rxidx} = (\x{\rxidx}, \y{\rxidx}, h) \in \mathbb{R}^{3}$.
These sampled RXs represent spatial evaluation points that approximate $\zoneInside$ and $\zoneOutside$, effectively realizing quasi-Monte Carlo integration~\cite{caflisch1998monte} over the target zone and its complement as defined in the objective of Section~\ref{sec:problem}.

\subsection{Channel Model}
\label{subsec:channel}

The channel from airborne node $\nodeidx \in \nodeset$ to the sampled RX location $\rxidx \in \mathcal{M}$ is modeled with wireless ray tracing ($\RT$), which resolves each link into a finite set of paths whose cardinality is determined by each node's configuration (e.g. position) and the geometry of the propagation environment. To form a coherent channel, we assume that each resolved path is geometrically feasible from $\nodeidx$ to $\rxidx$ and carries an associated time delay, power attenuation (loss), and phase.
Unique to RT is the ability to generate a \emph{deterministic} set of \emph{site-specific} multipath interactions including specular and diffuse reflection, refraction, and diffraction~\cite{sionnarttech}. 
The channel impulse response collects these combinations of rays, so the received power at sample $\mathbf{r}_m$ from node $n$ is obtained through the incoherent combining of path powers, i.e., 
\begin{equation}
    \textstyle
    \powerRx{n, m}(\txConfig_{n}) = \powerTx{n} \cdot \sum_{p} \left|a_{\pidx}(\txConfig_{n}, \rx{\rxidx})\right|^2,
  \label{eq:prx}
\end{equation}
where $a_{\pidx}(\txConfig_{n}, \rx{\rxidx})$ is the complex-valued path amplitude, whose value depends on the transmit antenna's location, and orientation ($\txConfig_{n}$), and the receiver's location ($\rx{\rxidx}$).
We adopt incoherent combining of all paths' energy, because it is insensitive to the sub-wavelength phase alignment, and therefore yields a smoother objective for gradient-based optimization.
{\eqref{eq:prx}} applies identically to both $\ABS$ and $\cjacro$s.

\subsection{Best-Server SINR}
\label{subsec:sinr}

For the $\rxidx$-th RX at $\rx{\rxidx}$, we assume a \emph{best-server} rule: the serving node is selected as the ABS of the highest SINR:
\begin{equation}
    P^{\star}_m(\txConfigSet) = \max_{n \in \txset} \left\{ P_{n, m}(\txConfig_{n}) \right\},
    \label{eq:bestserver}
\end{equation}
modeling an idealized handoff that reports the best achievable SINR at that location.
Since the CJ-induced interference and noise are common to all candidate serving links at $\mathbf{r}_m$, the resulting SINR is
\begin{align}
  \textstyle
  \SINR_{m} (\txConfigSet, \boldsymbol{\nu}) =
  \frac{P^{\star}_m(\txConfigSet)}
  {\sum\limits_{n \in \mathcal{B}}
    \powerRx{n,m}(\txConfig_{n}) - P^{\star}_m(\txConfigSet)
  + \sum\limits_{n \in \mathcal{J}} \nu_n \cdot \powerRx{n,m}(\txConfig_{n})
  + \powerNoise},
  \label{eq:sinr}
\end{align}
where $\boldsymbol{\nu} = [\nu_n]$ with $\nu_n \in \{0, 1\}$ being the activation gate of CJ $n \in \mathcal{J}$.
Specifically, in the denominator, we have (\emph{i}) the interference from non-serving ABSs, (\emph{ii})) the interference from CJs, and (\emph{iii}) $\powerNoise$={-100}\thinspace{dBm} as the noise power. 
Then, the corresponding SINR in dB-scale is given by
\begin{align}
    \SINR_{m, \text{dB}}(\txConfigSet, \boldsymbol{\nu}) = 10\log_{10} \left( \SINRlin(\txConfigSet, \boldsymbol{\nu}) \right).
    \label{eq:sinr_db}
\end{align}

%% file: sections/problem_formulation.tex
\section{Problem Formulation and Algorithm}
\label{sec:problem}

Tactical deployments rarely call for blanket coverage of a geometric region; they call for service that is \emph{confined to a specific, mission-defined area}.
{\system} considers an RF containment problem: given an operator-defined or application-specific zone, we have such a partition of $\zoneInside$ and $\zoneOutside$ that satisfies:
(\emph{i}) Across $\zoneInside$, the SINR must be driven \emph{high} so that intended users are served; whereas
(\emph{ii}) Across $\zoneOutside$, the SINR must be driven \emph{low}: exterior signal is leakage--it is wasted power, a source of interference to co-located systems, and the footprint over which the transmission is externally observable.
Minimizing SINR in $\zoneOutside$ shrinks that footprint without reference to any particular adversary, making the formulation \emph{target-agnostic}: we deny usable signal everywhere beyond $\zoneInside$ rather than nulling a known listener.

These two objectives map onto the two terms in the SINR {\eqref{eq:sinr}} and are served by two distinct node roles. The numerator is controlled by capable, fully optimized ABSs ($\txset$), which provision the desired signal and raise the SINR in $\zoneInside$. The denominator is controlled by CJs ($\jamset$), which raises the interference floor in $\zoneOutside$ and collapses the SINR therein. The two coupled and competing roles make the optimization challenging and intractable: a CJ placed to suppress exterior leakage also leaks interference inward, and a base station placed to serve the interior also leaks signal outward, so containment cannot be achieved by optimizing separately.

A natural question is why exterior suppression is delegated to cooperative jammers rather than additional service base stations. Structurally, lowering SINR requires an interference source, and a base station placed in {$\zoneOutside$} creates coverage rather than suppressing it. A noise-like jammer is preferable to a structured emitter for several reasons: it blends into the thermal floor and exposes no waveform structure or identity, alternative to traditional waveform jamming tactics presented and evaluated in \cite{hunt2023madradar}; it requires no coordination across platforms, unlike targeted nulling, which demands tight synchronization and precoding over limited backhaul; and it blankets a broad area regardless of shape, matching our need to trace a non-convex boundary.

\subsection{Optimization Problem}

{\system} jointly optimizes all airborne nodes configuration $\txConfigSet = \left\{ \txConfig_{\nodeidx} \right\}_{\nodeidx \in \nodeset}$, and a learnable parameter vector, 
$\boldsymbol{\Psi} = \left[ \psi_{\nodeidx} \right]_{\nodeidx \in \jamset}$, which determines which CJs should be activated.
The goal is to maximize SINR coverage across $\zoneInside$ while suppressing leakage across $\zoneOutside$, subject to the corresponding physical limits.
This optimization is formulated as:
\begin{tcolorbox}[
 boxsep=0mm,left=1mm,right=5.5mm,top=1.0mm,bottom=4mm,
 toptitle=1mm,bottomtitle=1mm,lefttitle=2mm,
 title={Optimization objective},
 boxrule=1pt,sharp corners=all]
\begin{align}
    \min_{\txConfigSet, \boldsymbol{\Psi}}:~
    &
    \Loss(\txConfigSet, \boldsymbol{\Psi})
    = \wIn\Lin(\txConfigSet, \boldsymbol{\Psi}) + \wOut\Lout(\txConfigSet, \boldsymbol{\Psi}) \nonumber \\
    & \qquad\qquad\quad + \wUnif\Lunif(\txConfigSet, \boldsymbol{\Psi}) + \wGate\Lgate(\boldsymbol{\Psi})
    \label{eq:opt-objective} \\
    \text{s.t.:}~
    &\label{cons:1} (\x{\nodeidx},\y{\nodeidx}) \in \zoneInside \cup \zoneOutside,\ \z{\nodeidx} \in (0, \z{\textrm{max}}), \\
    & \label{cons:2} \powerTxMin \le \powerTx{\nodeidx} \le \powerTxMax, \\
    & \label{cons:3} \az{\nodeidx}\in[0, 2\pi),\ \el{\nodeidx}\in[\elmin,\elmax],\ \forall \nodeidx\in\nodeset.
\end{align}
\end{tcolorbox}

\noindent
Here, we set the weights $\wIn=2.0$, $\wOut=8.0$, $\wUnif=0.1$, and $\wGate=3.0$ as the operator's relative priority among the four objectives. They are all evaluated by Monte Carlo integration across the discrete RX samples with their $\SINR_{m, \text{dB}}$ {\eqref{eq:sinr_db}}.

\myparatight{Maximizing inside coverage.}
This term is designed to improve the SINR performance within $\zoneInside$ from two perspectives.
First, it increases the SINR of locations whose received SINR falls below a target threshold $\SINRthresh$.
Second, it promotes a higher average SINR across $\zoneInside$.
Specifically, the loss is defined as
\begin{align}
\Lin(\txConfigSet, \boldsymbol{\Psi})
& \textstyle = - \Big[
 \frac{1}{|\mathcal{M}_{\text{in}}|}
  \sum\limits_{m \in \mathcal{M}_{\text{in}}}
  \splus^{2} (
        \widetilde{\SINR}_{m,\textrm{dB}}(\txConfigSet, \boldsymbol{\Psi}) - \SINRthresh
  )
\nonumber\\
& \textstyle \qquad +
  \frac{\muin}{|\rxSetInside|}
  \sum\limits_{m \in \rxSetInside}
  \widetilde{\SINR}_{m,\textrm{dB}}(\txConfigSet, \boldsymbol{\Psi})
\Big],
\label{eq:lin}
\end{align}
where $\muin=3.0$ is a weight that adds a priority on increases in the mean SINR in-zone. $\splus(x) = \ln(1+e^x)$ denotes the softplus function~\cite{dugas2000incorporating}. 
This function serves as a smooth approximation of the hinge/ReLU operator, providing non-zero gradients near the threshold and improving optimization stability. 
The first term, which is squared, places greater emphasis on locations whose SINR falls below $\SINRthresh$=0.0, while the second term maximizes the average SINR over $\zoneInside$.

\myparatight{Minimizing outside leakage.}
Similarly, the outside loss is designed to suppress the SINR in $\zoneOutside$ through two complementary objectives.
First, it strongly penalizes RX samples whose SINR exceeds the prescribed threshold $\SINRthresh$. 
Second, it further reduces the overall exterior SINR level by minimizing the average SINR across $\zoneOutside$. 
This loss is defined as 
\begin{align}
    \Lout(\txConfigSet, \boldsymbol{\Psi})
    & \textstyle = \frac{1}{|\mathcal{M}_{\text{out}}|}
    \sum\limits_{m \in \mathcal{M}_{\text{out}}}
    \splus^2 (
    \widetilde{\SINR}_{m,\textrm{dB}}(\txConfigSet, \boldsymbol{\Psi}) - \SINRthresh
    ) \nonumber \\
    & \textstyle \quad + \frac{\muout}{|\mathcal{M}_{\text{out}}|}
    \sum\limits_{m \in \mathcal{M}_{\text{out}}}
    \widetilde{\SINR}_{m,\textrm{dB}}(\txConfigSet, \boldsymbol{\Psi}),
\label{eq:lout}
\end{align}
where the first term penalizes only those exterior locations whose SINR exceeds the threshold $\SINRthresh$, while the second term drives down the average exterior SINR over the entire outside region; we also have the hyperparameter $\mu_{\text{out}}=1.0$ that controls the trade-off between these two objectives.

\myparatight{Ensuring uniform SINR distribution.}
The third loss is designed to reduce spatial fluctuations on SINR within $\zoneInside$ and to prevent the formation of low-SINR regions, or \emph{SINR holes}.
It consists of two terms: the first reduces the spatial variation of the SINR within $\zoneInside$, while the second penalizes locations whose SINR falls below a prescribed acceptable level.
Specifically, the uniformity loss is defined as
\begin{align}
    \Lunif(\txConfigSet, \boldsymbol{\Psi})
    & \textstyle  =
    \operatorname{Var}_{\rxSetInside} \left( \widetilde{\SINR}_{m, \text{dB}}(\txConfigSet, \boldsymbol{\Psi}) \right) \nonumber \\
    & \textstyle \quad + \frac{1}{|\rxSetInside|} \sum\limits_{m \in \rxSetInside} \big[ ( \SINRfloor-\widetilde{\SINR}_{m, \text{dB}}(\txConfigSet, \boldsymbol{\Psi}))^{+} \big]^2,
    \label{eq:lunif}
\end{align}
where $\operatorname{Var}_{\rxSetInside}[\cdot]$ denotes the variance across $\rxSetInside$, $\SINRfloor=10$\thinspace{dB} is the minimum acceptable $\SINR$ in $\zoneInside$, and $(x)^{+} = \max\{x,0\}$.

\myparatight{Gating CJ activities.}
To determine which CJs should be activated, we introduce a learnable parameter vector 
$\boldsymbol{\Psi} = \left[ \psi_{\nodeidx} \right]_{\nodeidx \in \jamset}, \psi_{\nodeidx} \in \mathbb{R}$.
During optimization, a sigmoid function is applied to each $\psi_{\nodeidx}$ to obtain a soft activity indicator,
\begin{align}
\textstyle
\widetilde{\nu}_{\nodeidx}(\psi_{\nodeidx}) = \left( 1+e^{-\psi_{\nodeidx}} \right)^{-1},
\end{align}
which maps the unconstrained $\psi_{\nodeidx}$ to the interval $(0,1)$.
The CJ gating loss is then defined as a function of $\boldsymbol{\Psi}$:
\begin{align}
\textstyle
    \Lgate(\boldsymbol{\Psi})
    = \frac{1}{|\jamset|} \sum\limits_{\nodeidx \in \jamset} \widetilde{\nu}_{\nodeidx}(\psi_{\nodeidx})
    = \frac{1}{|\jamset|} \sum\limits_{\nodeidx \in \jamset} \left(1+e^{-\psi_n}\right)^{-1},
\end{align}
which penalizes the average CJ activity and encourages sparse selection of candidate CJs.
Since minimizing $\Lgate(\boldsymbol{\Psi})$ drives the soft activity indicators toward zero, a CJ remains active only when its contribution to the overall objective {\eqref{eq:opt-objective}} outweighs the corresponding activity penalty.
Therefore, we can rewrite {\eqref{eq:sinr}} for a soft $\widetilde{\SINR}_{m}$ as a function of $\txConfigSet$ and $\boldsymbol{\Psi}$, which is
\begin{align}
  \textstyle \widetilde{\SINR}_{m} (\txConfigSet, \boldsymbol{\Psi}) = 
\frac{P^{\star}_m(\txConfigSet)}
  { \sum\limits_{n \in \mathcal{B}}
    \powerRx{n,m}(\txConfig_{n}) - P^{\star}_m(\txConfigSet)
  + \sum\limits_{n \in \mathcal{J}} \widetilde{\nu}_{\nodeidx}(\psi_{\nodeidx}) \cdot \powerRx{n,m}(\txConfig_{n})
  + \powerNoise}.
\label{eq:sinr-new}
\end{align}

After optimization, the final CJ selection is obtained by thresholding the optimized soft activity indicators, i.e.,
\begin{align}
\nu_{\nodeidx} =
    \begin{cases}
    1, & \widetilde{\nu} \ge \nu_{\text{th}}, \\
    0, & \text{otherwise},
    \end{cases}
    \label{eq:logit}
\end{align}
where $\nu_{\text{th}} = 0.5$ is a predefined threshold.
Here, $\nu_{\nodeidx} = 1$ indicates that CJ $\nodeidx$ is selected for activation, while $\nu_{\nodeidx} = 0$ indicates that it is deactivated. By plugging $\nu_{\nodeidx}$ in equation~\eqref{eq:sinr}, we can derive the final $\mathrm{SINR}_{m}$ for each receiver.

\begin{figure*}[t]
    \centering
    \vspace{-3mm}
    \includegraphics[width=.98\textwidth]{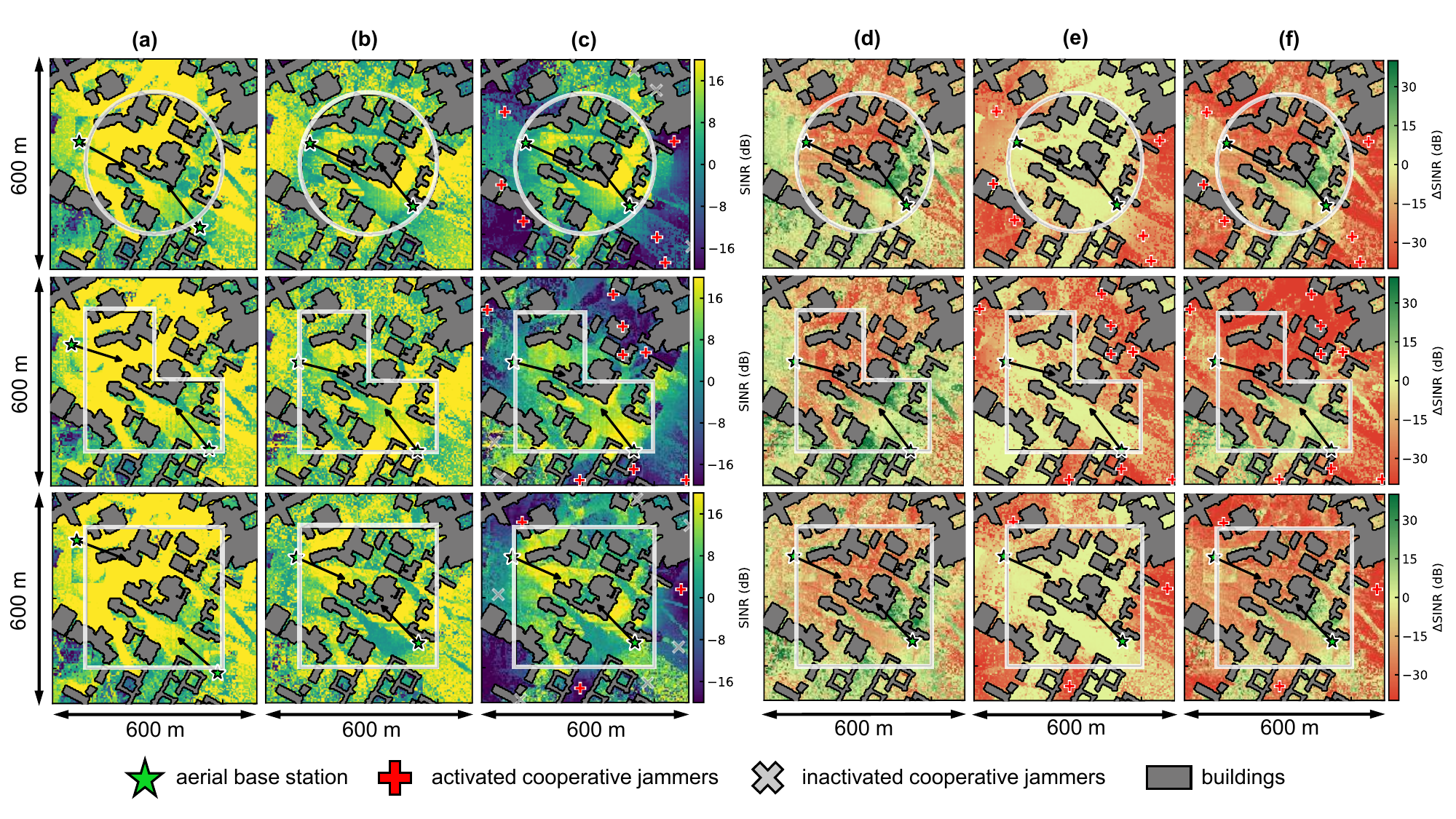}
    \caption{Locations and orientations of ABSs and CJs, and the resulting SINR distributions for the three zone geometries (Circle, L-shape, and Square): (a) initial ABS deployment without CJs; (b) optimized ABS deployment without CJs; (c) jointly optimized ABS and CJ deployment; (d) improved SINR uniformity by ABS optimization, i.e., the difference between (a) and (b); (e) exterior coverage degradation induced by CJ deployment, i.e., the difference between (b) and (c); (f) total shift in coverage between the optimized and initial configurations, i.e., the difference between (a) and (c).}
    \label{fig:heatmaps}
    \vspace{-3mm}
\end{figure*}

\subsection{The {\system} Algorithm}
 
{\system} seeds every node at a geometrically suitable location (i.e., no obstructions) before optimization, providing the first-order optimizer a favorable basin to solve for its achievable minima.
The node counts, $|\txset|$ for ABS and $|\jamset|$ for CJs, are determined by the number of available platforms.

\myparatight{Initializing ABS and CJ locations.}
To initialize the ABS locations, {\system} first generates a sufficiently large set of random candidate positions within $\zoneInside$ (e.g., 4,000), which are then clustered into $|\txset|$ groups using $k$-means++~\cite{kmeansplus}, producing $|\txset|$ representative centroids that are well distributed across $\zoneInside$.
Each centroid is then projected onto its nearest point on the boundary $\partial\zoneInside$ as an initial ABS location, whose azimuth angle is set to point toward the geometric center of $\zoneInside$.
This procedure provides a set of boundary-mounted TXs that remain spatially separated while collectively covering the interior of the zone.
The initial CJ locations are designed to surround $\areaTarget$ uniformly, increasing the likelihood of suppressing detectable leakage. 
{\system} first specifies a standoff distance, denoted by $L$, and offset the boundary of $\areaTarget$, $\partial \areaTarget$, outward by $L$ to obtain an expanded boundary. 
It then places the $|\jamset|$ candidate CJ at uniformly spaced locations along the expanded boundary using equal arc-length intervals around the perimeter.
This produces a ring of CJs that encircles $\zoneInside$ at a fixed separation from its boundary.

\myparatight{Two-Stage optimization.}
{\system} optimizes $\txConfigSet$ and $\boldsymbol{\Psi}$ using the Adam optimizer~\cite{kingma2017adammethodstochasticoptimization} for $2\iterstage$ iterations. The first stage (\emph{warm-start}) optimizes the ABSs alone with $\wOut = \wGate = 0$ and $\boldsymbol{\Psi} = \mathbf{0}$ {\eqref{eq:opt-objective}} for $\iterstage$ iterations, the latter is set to ensure that all the CJs are off.
This stage attempts to establish uniform interior coverage while minimizing $\Lin$ without competing with the leakage objective.
The second stage (\emph{full optimization}) starts with the ABSs from the warm-start solution, introduces the CJs and their activity gates, and proceeds with the full objective {\eqref{eq:opt-objective}} for $\iterstage$ iterations.
In this stage, $(\txConfigSet, \boldsymbol{\Psi})$ are jointly optimized so that the gates activate only the CJs that reduce exterior leakage while the interior coverage attained in the first stage is preserved.
$\txConfigSet$ and $\boldsymbol{\Psi}$ are optimized using different learning rates, each annealed by a cosine scheduling for termination at $\iter{}$.
All gradients are clipped before each step to guard against large log-gradients near the SINR floor. After optimization, we evaluate {\eqref{eq:logit}} to apply the hard CJ gates $\boldsymbol{\nu}$ from the soft gates $\widetilde{\boldsymbol{\nu}}$, and subsequently the SINR $\SINR$ from $\widetilde{\SINR}$.

%% file: sections/experiments.tex
\section{Evaluation}
\label{sec:experiments}

\subsection{3D Scene and Target Zone Generation}

{\system} generates the 3D model of an operational environment using Geo2SigMap~\cite{li2024geo2sigmaphighfidelityrfsignal}, a high-fidelity 3D scene generation pipeline that captures geographically-accurate environment geometries directly from latitude-longitude coordinates of a target area. Ultimately, the scene characteristics determine the per-interaction (reflection, diffraction, etc.) behavior and thus the fidelity of the downstream coverage predictions. Each scene is exported as a Mitsuba~3-compliant file referencing the triangle-mesh objects of the layout and imported into Sionna RT directly for evaluation using the refined spatial formulation.

The target zone $\areaTarget$ is constructed per scenario using Shapely, a library for planar geometric analysis~\cite{shapely_2025}.
The operator first approximates $\areaTarget$ as an ordered array of vertices instantiated as a Shapely object; a curved boundary is further approximated by densifying the number of vertices stored in the object.
The building contours reported by Geo2SigMap are likewise converted to polygon objects and unioned into $\areaBuilding$.
$\zoneInside$ and $\zoneOutside$ are then  constructed following Section~\ref{sec:system}.

Sionna~RT provides end-to-end differentiability of the traced ray paths based on Mitsuba~3's optical RT with PyTorch~\cite{paszke2019pytorch}, enabling the direct optimization of variables $\txConfigSet$ and $\boldsymbol{\Psi}$~\cite{sionnarttech}.
This lets {\system} cast RF containment as a first-order optimization problem over the downlink propagation from each airborne node $\nodeidx \in \nodeset$ (either ABS or CJ) to each sampled RX location $m \in \mathcal{M}$ (either in $\zoneInside$ or $\zoneOutside$).
All RT calculations utilize Sionna's point-to-point \texttt{\small{PathSolver}} class \cite{sionnarttech}.

\subsection{Monte Carlo Receiver Sampling}
\label{subsec:rxsampling}
The RX set $\rxSet$ contains the points at which the losses in {\eqref{eq:opt-objective}} are evaluated, so their placement and quantity determine the accuracy of the Monte Carlo estimates across $\zoneInside$ and $\zoneOutside$.
Each subset is independently and proportionally initialized using \emph{rejection sampling}, which distributes points using one sequence per coordinate dimension. {\system} uses a low-discrepancy sequence (Halton) to make the RX location sampling quasi-random, improving variance reduction in low-dimensional settings and thereby improving our approximations of $\Loss(\txConfigSet, \boldsymbol{\Psi})$ at each iteration~\cite{halton,caflisch1998monte}.

\subsection{Simulation Setup}

We evaluate {\system} using the \texttt{\small RadioMapSolver} class in Sionna RT, which generates wireless coverage maps over a 2D grid, producing spatially smoothed estimates of the Monte Carlo estimation described above. 
This method is analogous to estimating $\widetilde{\SINR}_{m,\text{dB}}(\txConfigSet, \boldsymbol{\Psi})$ using the quasi-randomly sampled point set $\rxSet$, but improves integration accuracy over $\areaTotal$ through a hardware-accelerated implementation of the shooting and bouncing of rays method~\cite{sionnarttech}.
Importantly, this solver class generates ray paths distinct from those of the \texttt{\small PathSolver} class, making it an unbiased evaluator.

In our evaluation, we consider the Duke University campus serviced by $|\mathcal{B}|=2$ ABSs and a candidate  set of $|\mathcal{J}|=12$ CJs. The number of in-zone receiver samples is configured to use $|\rxSetInside|=500$, while $|\rxSetOutside|$ is calculated proportionally based on area: $\bigl\lceil |\Sin|\,\operatorname{area}(\zoneOutside)/\operatorname{area}(\zoneInside) \bigr\rceil$. Our implementation presumes the use of all predefined optimization hyperparameters noted in \ref{sec:problem} and $\iterstage=100$ iterations per optimization stage.

Next, we define $\areaTotal$ using a {600}\thinspace{m}$\times${600}\thinspace{m} square, and partition it into a set of grid elements, each with a dimension of {0.5}\thinspace{m}$\times${0.5}\thinspace{m}, resulting in $1.44\times10^{6}$ grid elements.
We consider three geometries of the target zone $\areaTarget$: \emph{Circle}, \emph{L-shape}, and \emph{Square}; all of which have a max span from the zone center at $(0, 0)$ of {200}\thinspace{m} along the $x$- and $y$-axis.

\subsection{Impact of ABS and CJ Optimization on SINR Distribution}

Fig.~\ref{fig:heatmaps} illustrates the deployment locations of ABSs and CJs, together with the corresponding SINR distributions for the three zone geometries.
Fig.~\ref{fig:heatmaps}(a--b) compare the initial and optimized ABS deployments without CJs. For all geometries, optimizing the ABS locations alone degrades the SINR in the {\shapes} zones from {22.1/23.29/23.94}\thinspace{dB} to {14.9/18.0/16.3}\thinspace{dB}.
This is a consequence of the competing uniformity objective, as reflected by a subtle decrease in $\sigma$ from {13.1/15.6/14.2} to {11.9/13.1/12.5}\thinspace{dB}.
Outside the zone, the average SINR shifts from {18.3/20.6/17.5}\thinspace{dB} to {12.4/15.7/13.4}\thinspace{dB}, indicating that ABS optimization smoothed its interior signal coverage at the cost of increased leakage. Fig.~\ref{fig:heatmaps}(c) visualizes this difference directly.
Fig.~\ref{fig:heatmaps}(d) shows that jointly optimizing ABS and CJs substantially suppresses the mean SINR in $\zoneInside$ to {10.0/9.7/9.6}\thinspace{dB}, which closely matches the target $\SINRfloor$ value for all three versions of $\areaTarget$.
More notably, the mean SINRs of $\zoneOutside$ are degraded from {12.4/15.7/13.4}\thinspace{dB} to {$-$4.7/$-$3.1/$-$3.3}\thinspace{dB} using only {7/10/6} active CJs, confirming effective signal suppression achieved by {\system}.

\subsection{Coverage and Leakage Performance Analysis}

\begin{figure}[t]
    \centering
    \includegraphics[width=.98\columnwidth]{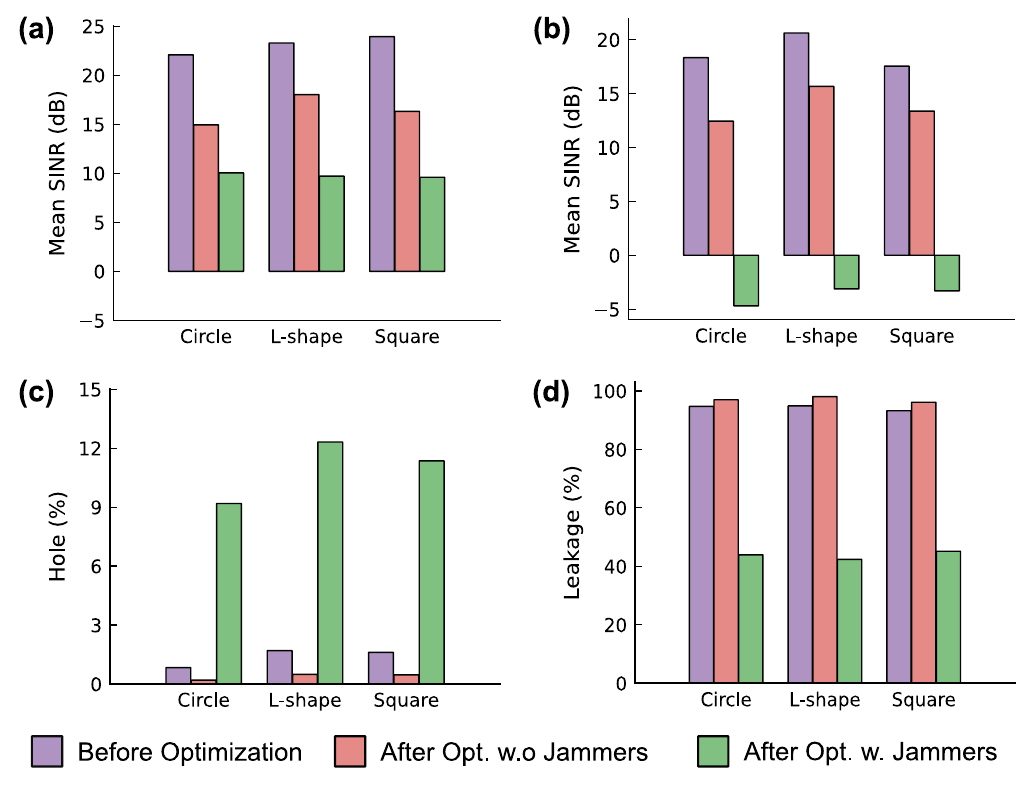}
    \vspace{-3mm}
    \caption{Performance on the three zone geometries (Circle, L-shape, and Square) before optimization, after ABS optimization without CJs, and after joint ABS--CJ optimization. (a) Mean SINR within the region $\zoneInside$; (b) Mean SINR within the region $\zoneOutside$; (c) percentage of $\zoneInside$ with SINR below $\gamma=${0}\thinspace{dB}; and (d) percentage of $\zoneOutside$ with SINR above $\gamma=${0}\thinspace{dB}.}
    \label{fig:duke-bar-final}
    \vspace{-3mm}
\end{figure}

Fig.~\ref{fig:duke-bar-final} quantitatively compares the performance of the three deployment strategies, namely the initial deployment, the optimized ABS deployment without CJs, and the joint ABS-CJ optimization.
Specifically, the mean SINR across $\zoneInside$ are shown in Fig.~\ref{fig:duke-bar-final}(a), and that across $\zoneOutside$ is shown in Fig.~\ref{fig:duke-bar-final}(b).
We further evaluate the effectiveness of RF containment using two complementary metrics computed over all grid elements in $\zoneInside \cup \zoneOutside$.
First, Fig.~\ref{fig:duke-bar-final}(c) reports the \emph{signal-hole ratio}, defined as the fraction of grid elements within the protected region whose SINR falls below the target threshold $\SINRthresh$.
This metric characterizes the reliability of service coverage inside the target zone.
{\system} exhibits signal-hole ratios of 9.2\%, 12.3\%, and 11.4\% for the circular, L-shaped, and square zones, respectively.
Second, Fig.~\ref{fig:duke-bar-final}(d) presents the \emph{signal-leakage ratio}, defined as the fraction of grid elements outside the protected region whose SINR exceeds $\SINRthresh$.
Specifically, the leakage ratios are reduced from 94.7\%, 94.9\%, 93.2\% pre-optimization to only 43.9\%, 42.4\%, and 45.2\% post-optimization for each of the three zones.
The modest signal-hole increases (9--12\%) reflect the expected tradeoff: CJs that suppress exterior leakage inevitably raise the interference floor near the zone boundary, slightly eroding coverage at a small number of interior edge cells while delivering large gains in containment outside.
Overall, these results demonstrate that {\system} effectively maintains an acceptable level of coverage within the target zone while suppressing signal exposure outside.

%% file: sections/conclusion.tex
\section{Conclusion}
\label{sec:conclusion}
We presented {\system}, a coverage-containment framework for aerial tactical networks. By jointly optimizing ABS and CJ deployment through differentiable ray tracing, {\system} confines wireless coverage to arbitrarily shaped, operator-defined target zones without any assumption on adversary location. Results presented in Section~\ref{sec:experiments} illustrate a significant impact on signal suppression and coverage retention across all three target zone geometries. These results confirm that denying usable signal beyond a mission-defined boundary, rather than concealing or attenuating the transmission itself, is both achievable and practical through gradient-based optimization against realistic propagation environments.

The present formulation assumes a static scene and a fixed set of candidate positions; extending \system{} to online re-optimization as nodes traverse dynamic or partially known environments is a natural next step. The current node model also treats each ABS as a single antenna; richer array architectures, e.g., analog multi-user beamforming with an array of subarrays~\cite{mambas2024}, could let one ABS serve multiple users within $\zoneInside$. As tactical networks continue to rely on rapidly deployable, ad hoc infrastructure, we view this containment approach as a foundation for denying adversaries usable signal in future contested environments.

%% file: sections/acknowledgements.tex
\section*{Acknowledgments}
\addcontentsline{toc}{section}{Acknowledgments}
This work was supported in part by NSF under Grants CNS-2211944, AST-2232458, ECCS-2434131, and CNS-2450567,
ARO under Award No. W911NF-25-1-0241,
AFOSR under Award No. FA9550-25-C-B010,
and NVIDIA Academic Grants. This material is based upon work supported by the Air Force Office of Scientific Research under award number FA9550-25-C-B010 in the amount of \$139,400.